# LOCKE Detailed Specification Tables

L.G. Menezo, V. Puente and J.-Á. Gregorio,

This document shows the detailed specification of LOCKE coherence protocol for each cache controller, using a table-based technique [1]. This representation provides clear, concise visual information yet includes sufficient detail (e.g., transient states) arguably lacking in the traditional, graphical form of state diagrams.

Table I shows detailed specifications of a L1 cache controller. States, events and actions are represented on it. Each of the rows of the table corresponds to a different possible state a cache line can be in. Each of the columns represents an event triggered by the coherence controller when receiving a message. The entries of the table show the actions that have to be taken on each case and sometimes the state the cache line has to change to (which is represented by '/' followed by the next state). For example: when a cache controller receives a *GETX* message from another and has the block requested in S state, it must send all the tokens to the requestor and jump to the *PS* state. If there is no state indicated, the cache line will stay in the same state after executing the corresponding actions.

The protocol has 5 stable states: *I*, *S*, *O*, *E* and *M*; it has 3 transient states: *IS*, *IM* and *SM*; and it has another 4 *control* states: *PS*, *PX*, *PO* and *F*.

Considering that the meaning of the stable and transient states mentioned is already known by the reader, we will explain the control states specific of LOCKE. The first three of them are states where the controller has sent some kind of data and is waiting for an acknowledgement message. The difference between them is the type of data sent. The *PS* state indicates that shared data was sent. *PX* state shows that data and all the tokens were sent (including the owner). *PO* state indicates that shared data was sent, but the controller still keeps the owner token. This differentiation is made to distinguish when the requests from others are needed to be answered. The last control state is the *frozen* state (*F*). A line gets to *F* when there is a pending write request to the address, but the controller has seen another write request (*GETX*) with more priority than it.

Almost all the times the events are triggered by the coherence controller depending on the messages received. There are some exceptions where the triggered event depends also on the state of the cache line. For example, when the processor issues a load or store operation, the cache controller triggers the load or store event if it has space for the new line. If it does not have free space, triggers a replacement event. Another example of this is the *GETX* and *FreezeGETX* events. The controller receives a *GETX* request from another controller but before triggering the event, checks if it has a pending *GETX* and whether its priority is greater or lower than the one received.

The *SpecialGETS* and *SpecialGETX* events are triggered when the controller receives a message of either of both types. These special messages are sent when a controller has a pending request and another controller informs of a possible tokens location. When received, they are treated as the normal *GETS* and *GETX* except for not being ignored when they cannot be solved.

Concerning to the ack event shown on the table, it is important to notice that the real controller needs to check if the ack received is the last one it is waiting for. If it is not, then the pending mark should be removed from the list of pending acks and stay in the same cache line state until the last ack is received.

The L2 controller, as Table II shows, works almost the same way as L1 controller does, with some exceptions. Obviously, there are no load or store operations to deal with. Also, as there is no store operation, *FreezeGETX* event disappears in this L2 controller. So do the retry and the complete event, as L2 cache will not receive retry or complete messages from other L1 caches.

Concerning to the L2 cache line states, there are also 5 stable states: *I, A, S, O, M*; and there are 4 control states: *PA, PT, PX, PO*. The new stable state A is needed to detect when a cache line has no valid data, but has tokens only. This situation happens when a L1 cache replaces a clean block and only sends tokens to L2. LOCKE control states in L2 are slightly different from the L1 controller, due to his particular behavior of always sending all the tokens available to any request. Therefore, *PA* state indicates that all tokens were sent and while expecting the acknowledgement, some replaced tokens from an L1 were received. *PT* indicates the same situation, but some data and tokens were received. *PX* and *PO* have the same meaning as in L1 controller.

**Table I State table for a L1 controller with LOCKE coherence protocol**

| | Load | Store | Replacement | Gets | Getx | FreezeGETX | SpecialGETS | SpecialGETX | DataShared | DataOwner | DataAllTokens | Ack | Retry | Complete |
|---|---|---|---|---|---|---|---|---|---|---|---|---|---|---|
| **I** | sendGETS | sendGETX | e | i | i | i | askToRetryBC | askToRetryBC | bounceData | bounceData | bounceData | e | i | i |
| **S** | do Load | sendGETX | replace /PS | i | sendAllToken /PS | sendAllToken /PX | askToRetryBC | sendAllToken / PS | update sendAck | update sendAck /O | update sendAck /M | e | i | i |
| **O** | do Load | sendGETX | replace /PX | send1Token /PO | sendAllTokens /PX | sendAllTokens /PX | send1Token /PO | sendAllTokens /PX | update sendAck | update sendAck | update sendAck /M | e | i | i |
| **E** | do Load | doStore | replace /PX | send1Token /PO | sendAllTokens /PX | sendAllTokens /PX | send1Token /PO | sendAllTokens /PX | e | e | e | e | i | i |
| **M** | do Load | doStore | replace /PX | send1Token /PO | sendAllTokens /PX | sendAllTokens /PX | send1Token /PO | sendAllTokens /PX | e | e | e | e | i | i |
| **IS** | z | z | z | i | i | i | askToRetryBC | askToRetryBC | update sendAck /S | update sendAck /O | update sendAck /M | e | sendSpecialGETS | i |
| **IM** | z | z | z | i | i | sendAllTokens /F | askToRetryBC | askToRetryBC | update sendAck /SM | update sendAck /SM | update sendAck /M | e | sendSpecialGETX | sendSpecialGETX |
| **SM** | z | z | z | askToRetryLater | i | sendAllTokens /F | askToRetryLater | askToRetryLater | update sendAck | update sendAck | update sendAck /M | e | sendSpecialGETX | sendSpecialGETX |
| **PS** | z | z | z | i | informTokenDest | informTokenDest | askRetryBC | informTokenDest | sendAck bounceL2 | sendAck bounceL2 /PX | sendAck bounceL2 /PX | /I | i | i |
| **PX** | z | z | z | informOwnerDest | informTokensDest | informTokensDest | informOwnerDest | informTokensDest | bounceL2 | bounceL2 | bounceL2 | /I | i | i |
| **PO** | z | z | z | send1Token | informTokensDest sendAllTokens /PX | informTokensDest sendAllTokens /PX | sendToken | informTokensDest sendAllTokens /PX | update sendAck | update sendAck | update sendAck | /I | i | i |
| **F** | z | z | z | retryWithBoss | retryWithBoss | retryWithBoss | i | i | bounceToBoss | bounceToBoss | bounceToBoss | /F | sendGETX | sendGETX |

Key: z → stall    i → ignore    e → error

<u>L1 States meaning</u>

- **I** : block not allocated
- **S** : data + token/s
- **O** : data + owner token
- **E** : clean data + all tokens
- **M** : modified data + all tokens

- **IS** : issued a GETS and waiting for data
- **IM** : issued a GETX and waiting for data
- **SM** : issued a GETS, waiting for data and received data with <u>some</u> of the tokens

- **PS** : data + token/s sent and waiting for acknowledge
- **PX** : data + all tokens sent and waiting for acknowledge
- **PO** : data + token sent, waiting for acknowledge and still have owner token
- **F** : frozen state – pending store operation but with less priority than another one

Table II. State table of a L2 controller with LOCKE coherence protocol

| | Replacement | L1_Gets | L1_Getx | SpecialGETS | SpecialGETX | DataShared | DataOwner | DataAllTokens | Tokens | Ack |
|---|---|---|---|---|---|---|---|---|---|---|
| I | e | i | i | askToRetryBC | askToRetryBC | storeData sendAck /S | storeData sendAck /O | storeData sendAck /M | updateNumTokens sendAck /A | e |
| A | issueWriteback /PX | i | sendTokens /PX | askToRetryBC | sendTokens /PX | storeData sendAck /S | storeData sendAck /O | storeData sendAck /M | updateNumTokens sendAck | e |
| S | issueWriteback /PX | i | sendAllToken /PX | askToRetryBC | sendAllToken / PX | updateNumTokens sendAck | updateNumTokens sendAck /O | updateNumTokens sendAck /M | updateNumTokens sendAck | e |
| O | issueWriteback /PX | send1Token /PO | sendAllTokens /PX | send1Token /PO | sendAllTokens /PX | e | updateNumTokens sendAck | updateNumTokens sendAck /M | updateNumTokens sendAck | e |
| M | issueWriteback /PX | sendAllTokens /PX | sendAllTokens /PX | sendAllTokens /PX | sendAllTokens /PX | e | e | e | e | e |
| PA | z | informOwnerDest | sendAllTokens informTokensDest / PX | askToRetryBC | sendAllTokens /PX | storeData sendAck /PT | storeData sendAck /PO | storeData sendAck /PO | updateNumTokens sendAck | /A |
| PT | z | informOwnerDest | sendAllTokens informTokenDest | askRetryBC | informTokenDest sendAllTokens | storeData sendAck | updateNumTokens sendAck /PO | updateNumTokens sendAck /PO | updateNumTokens sendAck | /S |
| PX | z | informOwnerDest | informTokensDest | informOwnerDest | informTokensDest | storeData sendAck /PT | storeData sendAck /PO | storeData sendAck /PO | updateNumTokens sendAck /PA | /I |
| PO | z | send1Token | informTokensDest sendAllTokens /PX | sendAllTokens /PX | informTokensDest sendAllTokens /PX | e | updateNumTokens sendAck | updateNumTokens sendAck | updateNumTokens sendAck | /I |

Key: z → stall   i → ignore   e → error

L2 States meaning

- I : block not allocated
- A : block allocated + only tokens (no data available)
- S : data + token/s
- O : data + owner token
- M : data + all tokens
- PA : data + all tokens sent, waiting for acknowledge and received some tokens
- PT : data + all tokens sent, waiting for acknowledge and received some data + token/s
- PX : data + all tokens sent and waiting for acknowledge
- PO : data + all tokens sent, waiting for acknowledge and received data + owner token